\documentclass[a4paper,11pt]{article}
\usepackage{pos}
\usepackage{bm}
\usepackage{booktabs}
\usepackage{tikz}
\usepackage{physics}
\usetikzlibrary{patterns}
\usepackage{pgfplots}
\usepackage{float}
\usepackage{subcaption}
\usepackage[utf8]{inputenc}
\DeclareUnicodeCharacter{2212}{-}
\DeclareUnicodeCharacter{22C6}{*}
\DeclareUnicodeCharacter{00A0}{α}
\pgfplotsset{compat=1.17}

\definecolor{myred}{RGB}{227,0,31}
\definecolor{myblue}{RGB}{56,125,216}

\title{Finite-Size Effects of the HVP Contribution to the Muon $g-2$ with C$^{\star}$ Boundary Conditions}
\ShortTitle{}

\author*[a]{Sofie Martins}
\author[b,c]{Agostino Patella}

\affiliation[a]{CP3-Origins \& IMADA, University of Southern Denmark, Campusvej 55, 5230 Odense M, Denmark}
\affiliation[b]{Humboldt Universität zu Berlin, Institut für Physik \& IRIS Adlershof, Zum Großen Windkanal 6, 12489 Berlin, Germany}
\affiliation[c]{DESY, Platanenallee 6, D-15738 Zeuthen, Germany}

\emailAdd{martinss@imada.sdu.dk}

\abstract{The muon $g-2$ is a compelling quantity due to the current standing tensions among the experimental average, data-driven theoretical results, and lattice results. Matching the final target accuracy of the experiments at Fermilab and J-PARC will constitute a major challenge for the lattice community in the coming years. For this reason, it is worthwhile to consider different options to keep the systematic errors under control. In this proceedings, we discuss finite-volume effects of the leading Hadron Vacuum Polarization (HVP) contribution to the muon $g-2$ in the presence of C$^{\star}$ boundary conditions. When considering isospin-breaking corrections to the HVP, C$^{\star}$ boundary conditions provide a possible consistent formulation of $\mathrm{QCD+QED}$ in finite volume. Even though these boundary conditions can be avoided in the calculation of the leading HVP contribution, we find the interesting result that they remove the leading exponential finite-volume correction. In practice, compared to the periodic case, C$^{\star}$ boundary conditions cut the finite-size effects in half on a lattice of physical size $M_{\pi}L=4$ and by a factor of almost ten for $M_{\pi}L=8$. We discuss the origin of this reduction and implications for computational efficiency.}

\FullConference{%
The 39th International Symposium on Lattice Field Theory,\\
8th-13th August, 2022,\\
Rheinische Friedrich-Wilhelms-Universität Bonn, Bonn, Germany
}

\begin{document}
\maketitle
	
\section{Motivation}
In this proceedings, we discuss finite-volume effects of the HVP contribution to the muon $g-2$ in isosymmetric QCD with C$^\star$ boundary conditions \citep{Kronfeld:1990qu, Kronfeld:1992ae, Polley:1990tf, wiese}. A priori, there is no reason to choose these boundary conditions for QCD observables. Nevertheless, it is interesting to consider this setup since we are interested in using it to calculate eventually also isospin-breaking corrections to the HVP. As discussed in \citep{Lucini:2015hfa}, C$^\star$ boundary conditions provide the only finite-volume formulation of QED which preserves locality, gauge invariance and translational invariance \citep{Lucini:2015hfa}. First numerical explorations in this direction have been presented in \cite{tavella_lattice, gruber_lattice, altherr_lattice}. We recall that locality can also be preserved by adding a mass term to the photon while breaking gauge invariance in a controlled way \citep{Clark:2022wjy, Endres:2015gda}.\par
Interestingly, we find that C$^{\star}$ boundary conditions reduce the finite-volume correction to the HVP contribution to the muon $g-2$. In fact, these are order $e^{-M_\pi L}$ and $e^{- \sqrt{2} M_\pi L}$ with periodic and C$^{\star}$ boundary conditions, respectively. While the proof of this statement will be given in a future publication, we sketch here the main ideas and the most notable differences with the periodic case, analyzed in \citep{Hansen:2019rbh, Hansen:2020whp}. We also provide an estimate of the finite-size corrections with C$^{\star}$ boundary conditions using a phenomenological model for the pion form factor.

\section{\texorpdfstring{C$^{\star}$}{} Boundary Conditions}
\paragraph{Definition.} \citep{Lucini:2015hfa} introduces the simulation of charged hadrons on the lattice using C$^{\star}$ boundary conditions. Quarks behave as in the following
\begin{equation}
	\Psi_{f}(x+L\hat{e}_{i})=\Psi_{f}(x)=C^{-1}\bar{\Psi}_{f}^{\mathrm{T}}(x)\,,
\end{equation}
\begin{equation}
	\bar{\Psi}_{f}(x+L\hat{e}_{i})=\bar{\Psi}_{f}(x)=-\Psi_{f}^{\mathrm{T}}(x)C\,,
\end{equation}
where $L$ is the lattice extent in space. Fields that are eigenstates of the charge conjugation operator behave as periodic or antiperiodic fields depending on their C-parity. For example, the electromagnetic current has negative C-parity and is an antiperiodic field in C$^{\star}$ boundary conditions:
\begin{equation}
j_{\mu}(x+L\hat{e}_{i})=j_{\mu}^{\mathrm{c}}(x)=-j_{\mu}(x)\,.
\end{equation}
As a consequence, the allowed momenta for the electromagnetic current are given by 
\begin{equation}
    \Pi_{-} = \left\{\dfrac{\pi}{L}(2\bm{n} + \bar{\bm{n}}) | \bm{n}\in\mathbb{Z}^3, \bar{\bm{n}} = (1,1,1)\right\}\,
\end{equation}
see \citep{Lucini:2015hfa}. This property has consequences for the definition of a finite-volume estimator for the leading-order HVP. While in the periodic case, one simply considers the zero-momentum two-point function of the electromagnetic current, this is not an option in the case of C$^\star$ boundary conditions, since the electromagnetic current does not have a zero-momentum component.
\paragraph{Finite-volume estimator.} Irrespectively of the boundary conditions, a finite-volume estimator of the leading-order HVP contribution to the muon $g-2$ can be defined as
\begin{equation}
a_{\mu}^{\mathrm{LO,HVP}}=\int_{0}^{T/2}\mathrm{d}x_{0}\,\mathcal{K}(x_{0})G(x_{0}|T,L) \,,
\end{equation}
where $\mathcal{K}(x_{0})$ is the kernel defined in \cite{Bernecker:2011gh}, and we define
\begin{equation}
	G(x_{0}|T,L)=-\dfrac{1}{3}\int_{V_{L}} \langle j_{k}(x)j_{k}(0)\rangle_{T,L}\mathrm{d}^{3}x\,.
\end{equation}
The integral of the two-point function is taken over a cube $V_L$ with an edge length equal to $L$. In the case of periodic boundary conditions, since the electromagnetic current is periodic in space, the integration domain $V_L$ can be freely translated without changing the value of the integral. In particular, the integral corresponds to a zero-momentum projection. In the case of C$^\star$ boundary conditions, since the electromagnetic current is antiperiodic, translating the integration domain $V_L$ will change the value of the integral, yielding a different finite-volume estimator. One can show that the symmetric choice
\begin{equation}
    V_{L}=\left(-\dfrac{L}{2},\dfrac{L}{2}\right)^3
\end{equation}
has the correct infinite-volume limit and yields finite-volume effects that are smaller with respect to the periodic case. Notice that, in the case of C$^\star$ boundary conditions, the integral over $V_L$ does not correspond to the projection over a single momentum: states propagating in between the two currents are superimpositions of all allowed momenta.\par
In principle, different finite-volume estimators can be constructed, e.g., by projecting the two-point function on one of the momenta allowed by the boundary conditions. For instance one could choose $\bm{p} = \frac{\pi}{L} (1,1,1).$ However, we found that this generates finite-volume corrections that vanish as inverse powers of $L$. For this reason, this choice will not be considered further.\par
\section{Finite-Size Effects in the Isospin-Symmetric Limit}
To derive an analytical formula for the finite-size effects for C$^{\star}$ boundary conditions, we can use the method developed in \citep{Hansen:2020whp} and apply it with modified boundary conditions. This calculation is non-trivial and beyond the scope of this proceedings. However, it is possible to examine the structure of the effects arising for C$^{\star}$ boundary conditions and compare them to periodic boundary conditions.
\paragraph{Necessary Adjustments.}
\citep{Hansen:2020whp} derives the finite-volume effects by expanding to all orders in an effective theory of pions. While in periodic boundary conditions, all pions are periodic fields, in C$^{\star}$ boundary conditions, their behavior is modified as in the following
\begin{equation}
\pi^{0}(x + L\hat{e}_i) = \pi^{0}(x)\,,
\end{equation}
\begin{equation}
\pi^{q}(x + L\hat{e}_i) = \pi^{-q}(x)\,.
\end{equation}
As a result, in contrast to periodic boundary conditions, we need to distinguish between contributions associated with different pion charges in the final formula . 
\paragraph{Analytic Results.}
Employing these adjustments, we find the following analytic expression
\begin{equation}
	\begin{split}\Delta G_{L}(x_{0})= & -\sum_{\bm{n}\neq\bm{0}}\sum_{q=\{0,\pm1\}}\chi_{q,\bm{n}}\int\dfrac{\mathrm{d}p_{3}}{2\pi}\dfrac{e^{-|\bm{n}|L\sqrt{M_{\pi}^{2}+p_{3}^{2}}}}{24\pi|\bm{n}|L}\int\dfrac{\mathrm{d}k_{3}}{2\pi}\:\mathrm{cos}(k_{3}x_{0})\mathrm{Re}T^{q}(-k_{3}^{2},-p_{3}k_{3})\\
		& \qquad\qquad\qquad\qquad\qquad\qquad\qquad\qquad\qquad\qquad+\mathcal{O}(e^{-\sqrt{2+\sqrt{3}}M_{\pi}L})
	\end{split}
 \label{result}
\end{equation}
where
\begin{equation}
	\chi_{q,\bm{n}}=\dfrac{1+(-1)^{q\sum_{i}n_{i}}}{2}\,,
\end{equation}
and $T^{q}(-k_{3}^{2},-p_{3}k_{3})$ is the forward Compton scattering amplitude of a pion with charge $q$ and momentum $(0,0,p_3)$ against an off-shell photon with Minkowskian four-momentum $(0,0,0,k_3)$, traced over the Lorenz-index of the photon. The adjustments due to the modified boundary conditions translate into this formula in the following way: Firstly, there is an explicit sum over different charges and, therefore, contributions that are specifically associated with the differently charged pions. For this, we need to calculate the Compton scattering amplitude of differently charged pions, here denoted $T^{q}$ instead of an implicitly summed pion Compton scattering amplitude $T$. \par
Secondly, the factor $\chi$ needs to be specifically introduced for C$^{\star}$ boundary conditions. 
For the contribution through the uncharged pion $\chi = 1$ for all $\bm{n}$, identically to periodic boundary conditions. \par
For the charged contributions, this factor can be either zero or one, depending on the vector $\bm{n}$. This vector can be interpreted as counting the number of times an internal pion line wraps around the torus, which is defined as $\sum_{i=1}^{3}n_i$. For $q\neq 0$, any odd wrapping number will yield $\chi=0$. As a result, contributions with odd $\bm{n}$ are filtered out, and only contributions for $q=0$ at the odd orders remain. This affects most notably the leading order. \par
\paragraph{Spectral Decomposition.} These uncharged contributions now turn out to be small already. This is because a spectral decomposition as in \citep{Hansen:2020whp} yields three contributions, a vacuum contribution, a one-particle pole contribution and remaining regular contributions.
\begin{equation}
T^{q} = T^{q}_{\mathrm{vac}} + T^{q}_{\mathrm{pole}} + T^{q}_{\mathrm{reg}}
\end{equation}\par
As in \citep{Hansen:2020whp}, we can assess all regular contributions, which is the regular part of the one-particle contribution and the multi-particle and higher-mass terms through chiral perturbation theory and find them to be small compared to all other effects. They are included in our numerical estimates.\par
The most significant part is the pole contribution through one-particle intermediate states. This term is proportional to the infinite-volume pion form factor and contributes to the finite-volume effects for the charged cases, which are removed through $\chi$ at odd orders. The remaining uncharged contributions are small because the form factor is zero for $q=0$.
\begin{equation}
T_{1\pi,\mathrm{pole}}^{0} \propto |\bra{\pi^{0}}J_\mu (0)\ket{\pi^{0}}|^2 = 0
\end{equation}
Further, the vacuum contribution is proportional to the matrix element
\begin{equation}
T_{\mathrm{vac}} \propto \bra{\pi^{q}} J_\mu\ket{\Omega}
\end{equation}
where $\ket{\Omega}$ denotes the vacuum. This is zero for charged pions thanks to charge conservation, and for the neutral pion thanks to C-parity conservation.\par

\paragraph{General Structure.} As for the periodic case discussed in \citep{Hansen:2019rbh} the effects exhibit exponential suppression with pion mass $M_{\pi}$ and lattice extent $L$, now with odd orders removed. Different orders of effects are generated by the sum over $\bm{n}$, generating terms of the structure
\begin{equation}
	\mathcal{O}(e^{-\sqrt{2}M_{\pi}L})+\mathcal{O}(e^{-2M_{\pi}L})+\mathcal{O}(e^{-\sqrt{6}M_{\pi}L})+\ldots\label{eq:finite-l-structure}
\end{equation}
followed by mixed finite-time and finite-volume effects
\begin{equation}
	\mathcal{O}(e^{-M_{\pi}T})+\ldots+\mathcal{O}(e^{-M_{\pi}\sqrt{L^2 + T^2}})+\ldots\label{eq:finite-t-structure}
\end{equation}
which for a setup of $T=2L$ vanish like $e^{-2M_{\pi}L}$. Finally, there are contributions through higher-mass hadrons
\begin{equation}
	\mathcal{O}(e^{-\sqrt{2}M_{K}L})+\mathcal{O}(e^{-2M_{K}L})+\ldots\label{eq:higher-mass-contrib-structure}
\end{equation}
which are also negligible because the next-higher-mass hadron is the kaon which has $M_{K}>2M_{\pi}$. Terms in \ref{eq:finite-t-structure} and \ref{eq:higher-mass-contrib-structure} are both neglected together with other contributions in the final formula for finite-volume effects at the order $\mathcal{O}(e^{-\sqrt{2+\sqrt{3}}M_{\pi}L})$.
\paragraph{Periodic Boundary Conditions.}
The main difference to the periodic boundary conditions from \citep{Hansen:2020whp} is the necessity to introduce the factor $\chi$. One can recover the analytic result for periodic boundary conditions by setting $\chi=1$ in eq.~\ref{result}.

\section{Results and Conclusion}
\begin{figure}
\caption{Finite-Volume Effects are cut in half for a physical lattice size of $M_{\pi}L=4$ when using C$^{\star}$ boundary conditions compared to periodic boundary conditions. For a large lattice of size $M_{\pi}L=8$ the effects are cut by a factor of almost ten. We are using a pion mass at the physical point of $M_{\pi}=137\,\mathrm{MeV}$. 
The finite-volume effects contribute at permil for C$^{\star}$ boundary conditions for a lattice of physical size $M_{\pi}L\approx 6$ while for periodic boundary conditions $M_{\pi}L\approx 8$ is necessary}
\centering
\begin{minipage}[t]{0.5\linewidth}
\centering
\subcaption{Absolute Finite-Volume Effects to the HVP}
\label{tbl:abs-fv}
\begin{tabular}{c|rr} 
$M_{\pi}L$ & C$^{\star}$ BC & PBC\cite{Hansen:2020whp} \\
\midrule
4 & 9.74(1.6) & 22.4(3.1)\\
5 & 3.25(0.23) & 10.0(0.4)\\
6 & 1.027(0.034) & 4.42(0.06)\\
7 & 0.311(0.005) & 1.924(0.009)\\
8 & 0.0909(0.0008) & 0.826(0.001)\\
\end{tabular}
\end{minipage}%
\begin{minipage}[t]{0.4\linewidth}
\centering
\subcaption{Relative Finite-Volume Effects to the HVP}
\label{tbl:rel-fv}
\begin{tabular}{c|rr}
$M_{\pi}L$ & C$^{\star}$ BC & PBC\cite{Hansen:2020whp} \\
\midrule
4 & 1.39 & 3.20\\
5 & 0.464 & 1.43\\
6 & 0.147 & 0.631\\
7 & 0.0444 & 0.275\\
8 & 0.0130 & 0.118\\
\end{tabular}
\end{minipage}
\end{figure}
\begin{figure}
\centering
\begin{tikzpicture}
\begin{axis}[
    title={},
    xlabel={Relative Contribution of FV Effects},
    ylabel={Estimated Cost Factor},
    xmode=log,
    x dir=reverse,
    xmin=0.005, xmax=1.5,
    ymin=-2, ymax=42,
    ytick={0,10,20,30,40},
    xtick={1, 0.1, 0.01},
    xticklabels={1\%, 0.1\%, 0.01\%},
    legend pos=north west,
    ymajorgrids=true,
    grid style=dashed,
]
\addplot[color=myblue,mark=*] coordinates {(1.39, 2.3)(0.46, 5.62)(0.15,11.64)(0.04,21.57)(0.01,36.80)}; 
\addplot[color=myred,mark=*] coordinates {(3.2,1)(1.43,2.44)(0.63,5.06)(0.27,9.38)(0.12,16)(0.05,25.63)(0.02,39.06)};
\legend{C$^{\star}$ BC,PBC};
\end{axis}
\end{tikzpicture}
\caption{Comparison of Finite-Size Effects Reached with Different Boundary Conditions}
\label{fig:prec-reached}
\end{figure}
\paragraph{Comparison of Finite-Volume Effects in Different Lattice Sizes.} We can evaluate these analytic formulae numerically analogously to the approach taken in \citep{Hansen:2020whp}. Table \ref{tbl:abs-fv} shows the absolute contributions of finite-volume effects to the HVP, comparing periodic boundary conditions (PBC) and C$^{\star}$ boundary conditions (C$^{\star}$ BC) for different physical lattice sizes. For a small lattice of $M_{\pi}L=4$, the finite-volume effects are cut by more than half compared to periodic boundary conditions. For a larger lattice of $M_{\pi}L=8$, this effect is even stronger: The effects are reduced by almost a factor of ten. From comparing the relative contributions in table \ref{tbl:rel-fv}, one can see permil finite-size effects are reached for periodic boundary conditions for $M_{\pi}L=8$ while for C$^{\star}$ boundary conditions, a lattice of physical size $M_{\pi}L=6$ is already sufficient.

\paragraph{Estimating Performance Differences.} When comparing the periodic and C$^\star$ setup, one needs to take into account the fact that simulations with C$^\star$ boundary conditions are generally more expensive. This is partly due to the fact that the lattice needs to be effectively doubled, and partly to the fact that the HMC algorithm for the light quarks needs to be replaced by the RHMC. We estimate that isosymmetric QCD simulations with C$^{\star}$ boundary conditions are about 2.3 times more expensive than simulations with periodic boundary conditions. Considering the computational cost, we can compare the merit of the different boundary conditions at different target finite-volume effects. Figure \ref{fig:prec-reached} shows an estimate of the computational time necessary to achieve a certain size in the finite-size effects comparing the two choices for boundary conditions. At a fixed relative size of the finite-volume effects, starting from 1\%, we expect C$^{\star}$ boundary conditions to outperform periodic boundary conditions consistently; however, for smaller target finite-volume effects, this improvement is marginal. Since we are targeting permil precision for the HVP, C$^{\star}$ boundary conditions are plausibly the better choice for evaluating the leading-order contribution.

\paragraph{Outlook.} In this proceedings we have described finite-volume corrections to the HVP contribution to the muon $g-2$, calculated in isosymmetric QCD with C$^\star$ boundary conditions. The proof of the formulae presented in this proceedings will be given in a future publication. The assessment of the finite-size effects in the isospin-breaking corrections to the HVP will be the subject of future work.

\paragraph{Acknowledgements.} S.M. received funding from the European Union's Horizon 2020 research and innovation program under the Marie Sk\l odowska-Curie grant agreement \textnumero~813942.

\bibliographystyle{apsrev4-1}
\bibliography{main}
\end{document}